\documentclass{article}

\usepackage{PRIMEarxiv}

\usepackage[utf8]{inputenc} 
\usepackage[T1]{fontenc}    
\usepackage{hyperref}       
\usepackage{url}            
\usepackage{booktabs}       
\usepackage{amsfonts}       
\usepackage{nicefrac}       
\usepackage{microtype}      
\usepackage{fancyhdr}       
\usepackage{graphicx}       
\usepackage{multirow}
\usepackage{amsmath}
\usepackage{wrapfig}
\usepackage{sidecap}

\usepackage{todonotes}

\title{Towards more efficient agricultural practices via transformer-based crop type classification
\thanks{ \textbf{Accepted at NeurIPS 2024 Workshop on Tackling Climate Change with Machine Learning}} 
}

\newcommand{\wrappedcell}[2][5cm]{%
  \begin{tabular}{@{}p{#1}@{}}%
  \parbox{#1}{\centering #2}%
  \end{tabular}%
}

\newcommand\jal{1}
\newcommand\mex{2}
\newcommand\ccai{3}
\newcommand\kab{4}
\newcommand\aspia{5}
\newcommand\colum{6}

\author{
\textbf{
  \wrappedcell[12cm]{
  Eduardo~Ulises~Moya~Sánchez$^{\jal,\mex}$,
  Yazid~S.~Mikail$^{\ccai}$,
  Daisy~Nyang'anyi$^{\kab}$, 
  Michael~J.~Smith$^{\aspia}$,
  Isabella~Smythe$^{\colum}$
  }}\\
  \vspace{0.01em}
  \\
  \wrappedcell[12cm]{
  $^{\jal}$Gobierno~del~estado~de~Jalisco,
  $^{\mex}$TecMM~Unidad~Zapopan,
  $^{\ccai}$Climate~Change~AI,
  $^{\kab}$Kabarak~University,
  $^{\aspia}$Aspia~Space,
  $^{\colum}$Columbia~University
  }
}

\begin{document}
\maketitle

\begin{abstract}
  Machine learning has great potential to increase crop production and resilience to climate change. Accurate maps of where crops are grown are a key input to a number of downstream policy and research applications.
  In this proposal, we present preliminary work showing that it is possible to accurately classify crops from time series derived from Sentinel 1 and 2 satellite imagery in Mexico using a pixel-based binary crop/non-crop time series transformer model. We also find preliminary evidence that meta-learning approaches supplemented with data from similar agro-ecological zones may improve model performance.
  Due to these promising results, we propose further development of this method with the goal of accurate multi-class crop classification in Jalisco, Mexico via meta-learning with a dataset comprising similar agro-ecological zones.
\end{abstract}


\section{Agriculture and AI}

Agriculture stands as the backbone of many economies worldwide, 
contributing as much as 60\% to a country's GDP and acting as a source of livelihood for over one billion people globally \cite{ref_worldbank2024,mbow2020food,musafiri2022adoption}.
The sector is highly affected by climate change due to the increasing prevalence of adverse weather conditions such as prolonged droughts, increased temperatures and precipitation, and unpredictable pests and diseases, all culminating in declining land productivity \cite{mbow2020food,musafiri2022adoption}.   
According to the FAO \cite{ref_fao2024}, about 757 million people worldwide face acute hunger in 2023. 
Moreover, the United Nations estimates a 25$\%$ global population growth by 2050, with developing and emerging economies within Latin America and Sub-Saharan Africa expected to grow the most \cite{falcon2022rethinking}.
This will ultimately put more pressure on already strained food systems, and we cannot simply clear more agricultural land; doing so results in deforestation and excessive use of resources such as environmentally damaging fertilizers and chemicals to increase productivity. 
Instead of brute-forcing a solution to this problem, we must find ways to use current resources more efficiently.


Accurate and timely crop classification is essential for centralized resource allocation, crop-specific yield prediction, and agricultural policy making.
For instance, allocating sufficient resources to farmers throughout the year allows planting flexibility and greater efficiency. 
Also, crop suitability to specific climates is crucial, as some crops require either cooler or warmer conditions, dictating optimal planting times. 
Additionally, local demand plays a significant role: even if a farmer successfully grows a large quantity of a product, it may not be possible to sell that product locally if there is insufficient demand or storage capacity.
Accurate and timely crop classification is therefore necessary for increasing land productivity. However, traditional crop classification methods like on-the-ground `field walks' and manual satellite imagery classification are error-prone and costly due to the need for a human-in-the-loop.
Machine learning methods do not have such downsides.
These automatic techniques can be deployed at scale far more efficiently than traditional crop classification as they do not rely on manual intervention once trained. 

This paper's contributions are two-fold: first, we present exploratory work on training a time series transformer classifier on satellite imagery time series to distinguish between agricultural and non-agricultural land. 
Second, we propose expanding this method to multi-class crop classification, addressing a critical data gap for policymakers in Jalisco, Mexico. %

\section{Data and methods}

\begin{wrapfigure}{r}{0.55\textwidth}
    \centering
    \includegraphics[width=1\linewidth]{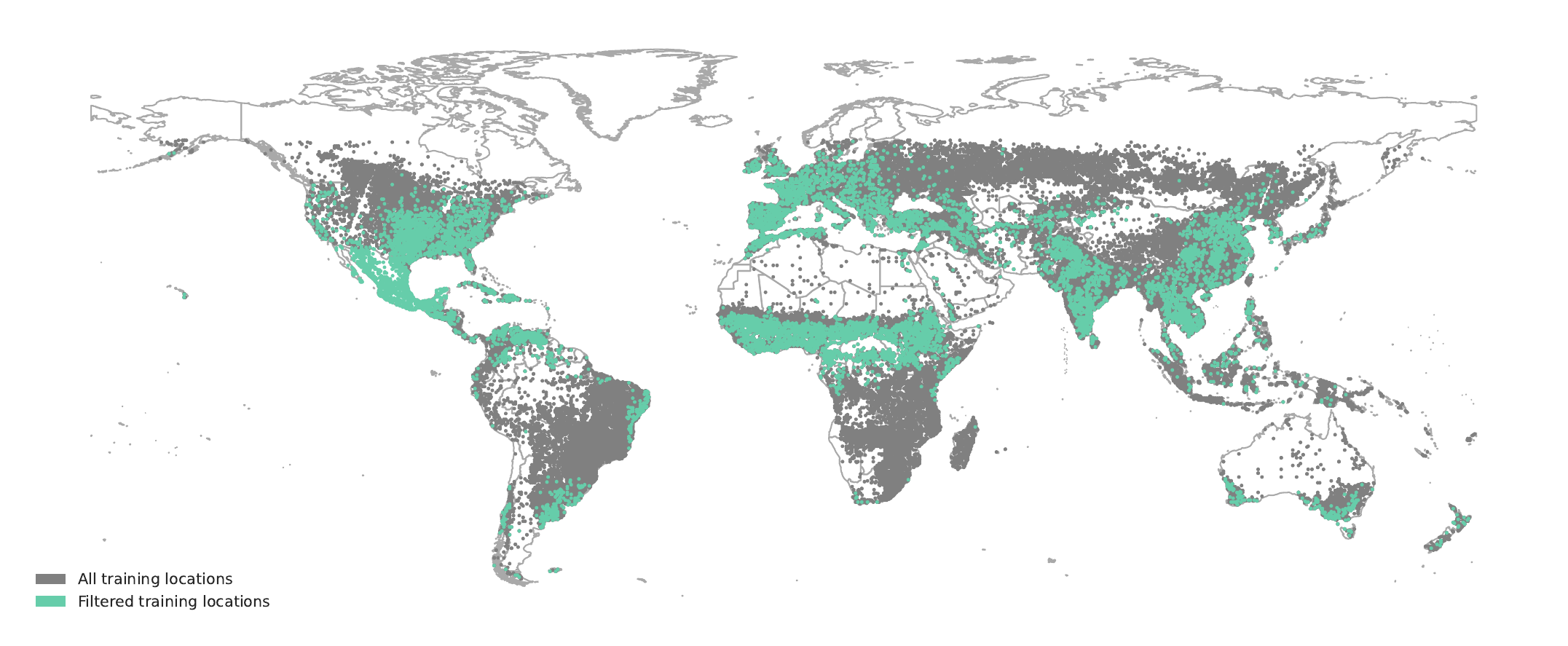}
    \caption{Locations of ground truth binary training labels are shown in grey. Locations included in the filtered sample of data with similar satellite time series to Mexico are shown in green.}
    \label{fig:locs}
\end{wrapfigure}

\begin{figure}[b]
    \centering
    \includegraphics[width=1\linewidth]{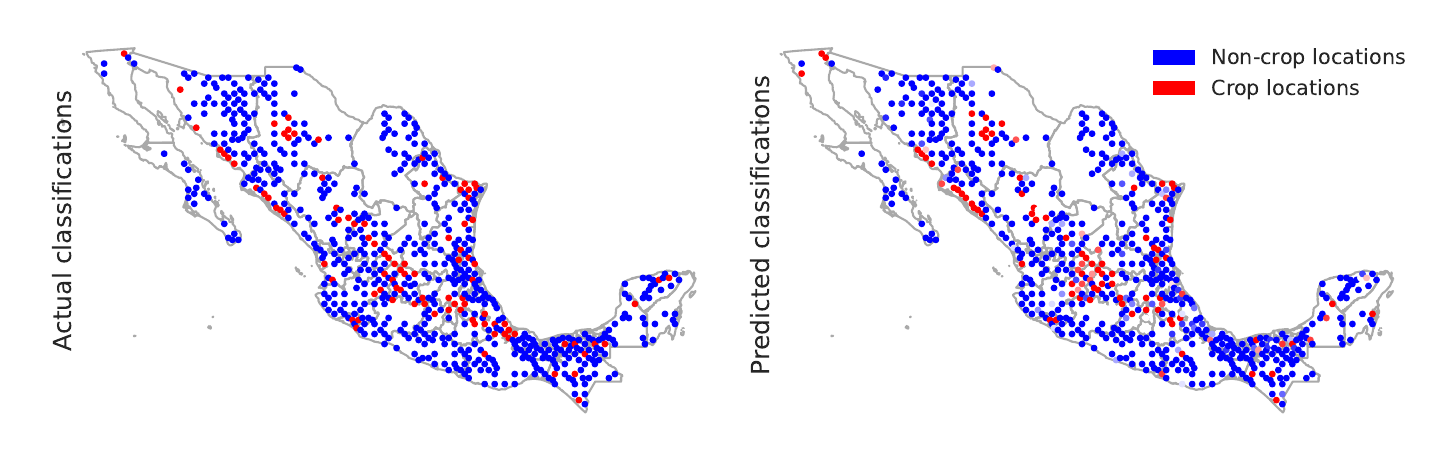}
    \caption{Actual versus predicted crop classifications. The left plot shows ground truth labels; the right plot shows labels predicted by a transformer model trained on the global dataset.}
    \label{fig:results}
\end{figure}

Our proof-of-concept work uses a simplified satellite imagery time series dataset, each corresponding to a location with a binary crop/non-crop label.
Upstream data sources are the NASA CropHarvest dataset \cite{OpenMapFlow2023} and labels from \href{https://geo-wiki.org/Application/}{Geo-wiki Landcover 2017}.
CropHarvest contains 12 Sentinel-2 bands, two Sentinel-1 bands, ERA5 monthly average temperature and precipitation, and SRTM DEM data on elevation and topography.
Sentinel data is processed to contain the least cloudy observation per month, with other observations discarded.
The model input is a 12-month yearly sample from February to February; all available data is for the 2017 growing season.

We use a three-layer transformer \cite{vaswani2017} model based on upstream code from~\cite{OpenMapFlow2023}.
Due to their self-attention mechanism, transformers allow efficient modeling of long-range dependencies in sequential data, making them well-suited for time series inputs like the ones used in this analysis.
The model is trained on three different datasets: Mexico data only, all global data, and a filtered subset of the international data intended to capture regions agroecologically similar to Mexico. There are several potential strategies for generating this filtered subset; as an initial approach, we assess similarity based on the satellite, weather, and topology data in the NASA CropHarvest dataset. Using Principal Component (PC) Analysis \cite{ref_karl1901}, the first 5 PCs for each pixel are calculated. For each PC, the central 90$\%$ of values found in the Mexico data constrains the allowed values; locations for which any PCs fall outside this range are removed from the training dataset.

For each sample, we perform $k$-fold cross-validation with $k=20$. Training data is sampled with replacement in each fold, and predictions are generated for the subset of Mexico data not used during training. 
Since the data spans only one year, training folds are assigned for Mexico at the state level\footnote{Mexico is divided into 31 states, the largest administrative subdivision.} to decrease overfitting due to spatial auto-correlation. Each pixel is assigned the label predicted in the majority of $k$-folds.

The default model hyperparameters from \cite{OpenMapFlow2023} are used for this preliminary analysis, but we expect improvements from implementing cross-validated parameter selection. For the global and filtered subset training samples, we use a learning rate of $10^{-4}$ and train for 25 epochs; for the Mexico-only model, a smaller learning rate and a larger number of epochs is used to improve stability ($\text{lr}=10^{-5}$ and 100 epochs). We use the Adam optimizer \cite{ref_kingma2014}.

\section{Results}
\begin{wraptable}{r}{0.55\textwidth}
    \vspace{1em}
    \caption{Results for proof-of-concept binary crop classifier. Column 2 shows the number of training samples; columns 3-5 show performance for a model trained on each region and tested in Mexico.
    }
    \centering
    \begin{tabular}{lcccc}
        \toprule
       \textbf{Region}   & $\mathbf{n}$ & \textbf{Acc.} & \textbf{Precision} & \textbf{Recall} \\
       \midrule
         Mexico  &  673 & 86.3 & 52.2 & 60.8 \\
         Global  & 34270 & 90.0 & 65.0 &  67.0 \\
         Filtered & 7656 & 88.0 & 56.9 & 68.0 \\
         \bottomrule
    \end{tabular}
    \label{tab:results}
\end{wraptable}

Results of preliminary models are shown in Table~\ref{tab:results}.
We find that expanding the pool of training data increases overall accuracy. 
However, the performance difference between models is relatively small, especially since default hyperparameters may be more appropriate for some training samples than others. The model's results trained on the filtered sample are promising, but additional work on filtering and meta-learning strategies is needed to determine whether this outperforms simpler approaches.
Figure~\ref{fig:results} compares ground truth crop locations and model predictions. Good general agreement corroborates the quantitative results in Table~\ref{tab:results}.

\section{Conclusions and proposal of future work}
The initial proof of concept implemented here shows promising accuracy for distinguishing crop- and non-crop pixels. Below, we highlight three significant directions for future work. 

First, this preliminary analysis involved binary, pixel-level classification on a simplified dataset. While improved accuracy for this task in historically understudied regions is valuable, the primary end goal of this work is multi-class segmentation on the full Jalisco dataset provided by the Jalisco government. Shifting to this more complex task will require adapting models from binary to multi-class estimation and combining classification with image segmentation to detect field boundaries.

Second, these preliminary results use a standard transformer model with little tuning or adaptation \cite{OpenMapFlow2023}. Developing task-specific transformer models and implementing cross-validated hyperparameter selection are likely necessary to obtain optimal performance.

Finally, we propose a more in-depth analysis of the use of dataset augmentation and meta-learning. In particular, the multi-class, Jalisco-specific dataset includes many more training data locations ($n = 35812$), which may reduce the benefit of these approaches. 
Consolidating the multi-class Jalisco labels with the binary supplementary data will also be necessary; for example, we can experiment with a pre-training binary classification step followed by a secondary model to assign crop type to regions classified as cropland. 
More robust experimentation with filtering approaches is needed, including hand-filtering locations as a baseline, better accounting for differences in the timing of the growing season between locations, and using variables that more directly capture relevant agroecological conditions. 
Additionally, we expect benefits from shifting to a meta-learning approach rather than the dataset augmentation used here.

\section*{Acknowledgments}

We thank Climate Change AI and Mila for hosting us during part of this research, and express our gratitude to the organisers of the 2024 Climate Change AI Summer School.

\bibliographystyle{unsrt}  
\bibliography{references}

\end{document}